\documentclass[aps,pra,twocolumn,groupedaddress,showpacs,superscriptaddress,scrartcl,longbibliography]{revtex4-2}
\usepackage{bbold,array,xcolor,enumitem,tabularx,multirow,hyperref,comment,amssymb,amsmath,graphicx,epstopdf,amsthm,dsfont,bm,color}

\newtheorem*{lemma1*}{Lemma 1}
\newtheorem*{lemma2*}{Lemma 2}
\newtheorem*{lemma3*}{Lemma 3}
\newtheorem*{lemma4*}{Lemma 4}
\newtheorem*{lemma5*}{Lemma 5}
\newtheorem*{lemma8*}{Lemma 8}
\newtheorem*{lemma9*}{Lemma 9}
\newtheorem*{lemma10*}{Lemma 10}
\newtheorem*{lemma11*}{Lemma 11}
\newtheorem*{theorem1*}{Theorem 1}

\newtheorem*{theorem2*}{Theorem 2}
\newtheorem*{theorem3*}{Theorem 3}

\newcommand{\edit}[1]{{\color{black}{#1}}}

\begin{document}

\title{\edit{Towards \edit{practical} quantum position \edit{verification}}}

\author{George Cowperthwaite}
\affiliation{Centre for Quantum Information and Foundations, DAMTP, Centre for Mathematical Sciences, University of Cambridge, Wilberforce Road, Cambridge, CB3 0WA, U.K.}

\author{Adrian Kent}
\email{apak@cam.ac.uk}
\affiliation{Centre for Quantum Information and Foundations, DAMTP, Centre for Mathematical Sciences, University of Cambridge, Wilberforce Road, Cambridge, CB3 0WA, U.K.}
\affiliation{Perimeter Institute for Theoretical
	Physics, 31 Caroline Street North, Waterloo, ON N2L 2Y5, Canada.}

      \author{Dami\'an Pital\'ua-Garc\'ia}
\email{D.Pitalua-Garcia@damtp.cam.ac.uk}
\affiliation{Centre for Quantum Information and Foundations, DAMTP, Centre for Mathematical Sciences, University of Cambridge, Wilberforce Road, Cambridge, CB3 0WA, U.K.}

\date{\edit{\today}}

\begin{abstract}

We discuss \edit{protocols for quantum position} verification schemes based on the
\edit{standard quantum cryptographic} assumption that a tagging device can keep classical data secure
\cite{K11.1}\edit{Our schemes use a classical key replenished by quantum key distribution. 
The position verification requires no quantum communication or quantum information processing.
The security of classical data makes the schemes secure against non-local spoofing attacks that 
apply to schemes that do not use secure tags.
The schemes are practical with current technology and allow for errors and losses.
We describe how a proof-of-principle demonstration might be carried out.} 
\end{abstract}

\maketitle

\section{Introduction}

The task of quantum position verification or authentication (QPV or QPA), also called quantum tagging, was first discussed in a patent \cite{patent} published in 2006. In the idealized version, the key idea is to challenge a tag (or prover) to
authenticate its location by sending it quantum and classical
communications at light speed, with instructions to process the
signals instantaneously and return responses at light speed. If the challenges are sent from several appropriately located test stations (or verifiers), relativistic signalling constraints can ensure that if the tag functions correctly,
but at a different location from that expected, this will
be detected by the verifiers, because there will necessarily
either be time delays or incorrect or missing responses.
Realistically, the signals should be sent as close as possible
to light speed, the processing should be as fast as possible,
and the protocol aims to guarantee location within as
small a region as possible. 

As spoofers cannot copy quantum information chosen from a non-classical ensemble (for example qubits in BB84 states), quantum tagging schemes are not generally vulnerable to the simple copy-and-redirect spoofing attacks applicable
to purely classical schemes. However, they may be vulnerable to
other attacks, as we now review.  

QPV was first discussed in the academic literature in
Refs. \cite{malaney2010location,chandran2009position},
which proposed schemes that were claimed to be unconditionally
secure. However, as \edit{first pointed out in Ref. \cite{KMS11}}, although the schemes in Refs. \cite{patent,malaney2010location,chandran2009position}
indeed protect against simple attacks that (only) copy and
broadcast classical signals and reroute quantum signals,
they are vulnerable to teleportation attacks, which effectively
simulate the operation of the tag at distant sites. Hence none of them are unconditionally secure.

More generally, Ref. \cite{buhrman2014position} showed that, for any scheme in this general class, the operations carried out by the tag can in principle be simulated (for the verifiers) by spoofers who are located between the tag and verifiers. These spoofing attacks involve non-local quantum computations using pre-distributed quantum entanglement. Since the known attacks require large amounts of pre-distributed entanglement and error-corrected quantum computation, the schemes may guarantee security given presently very credible technological assumptions. Nonetheless, they are not unconditionally secure.  

The no-go theorem of Ref. \cite{buhrman2014position} is a theoretically beautiful result, one of \edit{several 
(im)possibility theorems} in relativistic quantum information processing \edit({e.g., \cite{K99,K99.2,K13,PG15.1,hayden2016summoning,adlam2016quantum,kent2018unconstrained,PGK18,PG19, PG21})
that are either based on, or act as counterpoints to, fundamental results in non-relativistic quantum information processing 
(e.g. \cite{wootters1982single,dieks1982communication,mayers1997unconditionally,lo1997quantum,lo1998quantum,L97,gottesman2000theory,V03}).}  

However, in many (perhaps most?) plausible scenarios the possibility of quantum teleportation and non-local computation attacks is either unnecessary or insufficient to establish that practical quantum position verification is necessarily insecure. In essence, these non-local attacks establish that spoofers can deceive distant verifiers into believing that a tagging device is present at the expected location, when in fact it is not. But there is an important prior question: do verifiers generally really care
about the location of a tagging device {\it per se}? A quantum tag is essentially a small device that measures and perhaps applies unitaries to incoming quantum states, according to classical instructions, following a fixed public algorithm. As the name suggests, its purported role is to ensure the location of a tagged object or person, and {\it this} is what the verifiers care about.

In one scenario commonly considered in the literature, a position verification protocol is defined to allow a prover to prove their location \edit{($L$)} to distant verifiers. The protocol is said to be insecure if spoofers at other locations (not including \edit{$L$}) can simulate the actions of an honest prover when the prover is in fact absent. This is an important scenario, which captures the type of insecurity established by the results of Refs. \cite{KMS11,buhrman2014position}.

However it is also important to keep in mind that there are other interesting scenarios.  For example, the prover might be mislocated but still active, with spoofers using the prover's actions as part of their spoofing strategy in order to persuade the verifiers that the prover remains at  \edit{$L$}. The prover here might be oblivious to their mislocation (having wandered or been unwittingly deceived about their location), or might be cooperating with the spoofers (if, for example, they are a tagged prisoner attempting to escape
confinement with outside help).

It is also crucial to note that the term ``prover'' is ambiguous and potentially misleading when considering practical applications. As used in theoretical analyses, it generally represents the actions of a proving device (i.e. some
form of tag). However, the term suggests a person or agent. If \edit{Bob} carries a mobile device to respond to challenges from \edit{Alice's} various verifying stations, but \edit{he} and the device are separable, then the protocol may verify the device's location but not \edit{his}.

Moreover, to the extent that QPV protocols in the prover-verifier model are secure (given suitable technological bounds on the spoofers), they only establish the location of {\it a} suitable tagging device, not necessarily {\it \edit{Bob's}} tagging device. Any substitute device with the same functionality can implement the protocol correctly. We emphasize that, in the scenario under discussion, the spoofers know every detail of the tagging device's operations: the teleportation attacks of Ref. \cite{KMS11} and the more general non-local computation attacks of Ref. \cite{buhrman2014position} require this. So, if these attacks apply, constructing an identical tag is indeed an option for the spoofers.

Now, if spoofers want to deceive verifiers that someone or something is somewhere they are not, it is presumably either because the object or person has been destroyed or has been moved, or encouraged or deceived (perhaps by GPS spoofing) into moving, to the wrong location. To achieve this, one option is to detach the tag and leave it in position, while
dislocating the taggee. Another is to dislocate or destroy both tag and taggee, leaving an identical replica tag in the required position. Either way, the verifiers are left 
with the false impression that the tagged object remains in place, while in fact they have been untagged and/or dislocated and/or destroyed.   

It is true that, if the spoofers dislocate a tag but leave it
operational, it may continue to receive and respond to at least some of the verifiers' signals, and this might alert them to interference. Alternatively, some alarm device on the tagged object might alert the verifiers by signalling to them, or the tagged person might do so (in scenarios where they are cooperating with the verifiers). But note that these are also issues if spoofers apply teleportation or non-local computation attacks. To apply these, the spoofers must intercept (at least) quantum signals from the verifiers and use these to generate their spoofed responses. The prover will thus either not receive the quantum information components
of their challenges (in which case they might send an alarm signal) or else spoofed quantum information components generated by the spoofers. These spoofed quantum information components may be drawn from the same distribution as the original challenges but will not be identical to them.
So, the prover's responses will not generally be valid for the original challenges. If these responses reach the verifiers, they will, again be alerted.

However, in both cases, the spoofers can in principle prevent any alarm or false responses reaching the verifiers by screening the first tag and object/person, jamming all possible
communications between them and the verifiers. We emphasize, though, that they also need to do this if they apply teleportation \cite{KMS11} or non-local computation \cite{buhrman2014position} attacks, if the tag/object/person
is dislocated but not destroyed. So the need for jamming does not differentiate between these attacks and pure dislocation attacks.

There are conceivable scenarios that do differentiate between these attacks. For example, the prover could be in a region that the spoofers can surround, and signal into, but not enter.
The spoofers may be able to spoof GPS within the region, causing the prover to move to the wrong location. They may also be able to spoof the prover's responses to the verifiers using teleportation or non-local computation attacks, giving the verifiers the impression that the prover is at the correct location.  And they may be able to jam any communication from the prover to the verifiers, so that the verifiers receive only the spoofed responses. However, the spoofers may not be able to apply direct attacks involving physical dislocation or
detachment. The security assumptions here are somewhat delicate, because the spoofers must be able to send physical signals into the region in order to spoof the prover's GPS, but the power of anything they send into the region must be limited. They might perhaps be assumed to be able to send low-energy microwave and radio signals into the region but not, for instance, robots. This scenario requires quite strong and specific physical limitations on the spoofers, and so evidently does not allow unconditional cryptographic security.

More generally, precluding object dislocation/destruction and tag detachment/substitution attacks inevitably requires physical assumptions. For example, it must be hard to detach, destroy or move the tag quickly enough that the tagging protocol can continue with a substitute tag without an interruption being evident. The light speed signalling bound does give an unconditional constraint on movement. For example, a tag responding every microsecond could be moved no more than $\approx 300$m between responses. However, the signalling bound gives no effective unconditional constraint on attacks involving destruction and replacement: in principle,
a tag could be progressively destroyed on one side and replaced on the other side of a boundary that moves at near light speed, without detection. Defences against such attacks must be based on physical assumptions that cannot be unconditionally guaranteed. Also, in many scenarios, the light speed movement bound may be too weak, and stronger bounds again rely on further physical assumptions (for example, that the tag cannot be moved faster than the speed of sound, or perhaps than the fastest aeroplane developed to date).

In summary, security for any form of quantum tagging or position verification has to be based on some physical assumptions. We emphasize that this is not only because of non-local computation attacks, which can only be precluded by technological assumptions bounding the power of spoofers.
Dislocation/destruction and detachment/substitution attacks can
also only be precluded by physical assumptions on the properties of the tag (and its attachment).

Given that we need such assumptions anyway, there are strong
reasons for exploring tagging schemes based on a physical
assumption that is standard in classical and quantum cryptography, namely that a piece of infrastructure (in this case the tag) is able to store secret classical data.
Ref. \cite{K11.1} introduced tagging schemes that are unconditionally secure, modulo this assumption, in the sense that the tag itself cannot (except with small probability) be spoofed or replaced so long as it remains intact and the data it contains remains secret. In this paper, we explore further versions of this scheme and discuss their practical implementation.

\section{Security  scenario}  
 \label{secII}
 
We assume that the verifiers, referred to collectively as Alice and individually as  $A_1,\ldots, A_M$, and the prover (Bob), are human or other agents, all of whom trust one
another. They have a fixed agreed reference frame $F$.
For simplicity, we assume that Bob is supposed to remain at a constant location during the position verification scheme.
We assume that Alice has some prior unreliable knowledge
of Bob's location (e.g. a reported GPS reading from Bob, or knowledge of a pre-agreed location which Bob is supposed to reach), and wishes to verify it to within as small an error margin as possible, at some sequence of times $T_j$, for $j = 1
, \ldots , N$.   We assume that the $A_i$ can communicate
with each other via authenticated channels.   

In this simple scenario, the aim of the scheme is that, if Bob does remain at a fixed location $L$ during the scheme, he can 
allow Alice to verify that he is close to $L$ at the given times. In more general scenarios, Bob may move (perhaps up to some speed bound), and the aim may be to identify his position (up to some uncertainty) at any given time.  The protocols we describe can be adapted to these more general scenarios, but for clarity we focus here on the simple scenario. 

Position verification is non-trivial because of the potential presence of adversaries, who may interfere with, jam, or substitute signals between Alice and Bob.   They may also apply physical attacks -- for example physically dislocating Bob.  

We assume spacetime is Minkowski, as is approximately the case near the Earth surface. The small general relativistic corrections required can easily be allowed for if needed (i.e. if they are significant enough), but we ignore them here.
Our discussions assume an asymmetry between Alice, whose agents $A_i$ each are within and control separate secure laboratories, which we might imagine as well-resourced fixed bases, and Bob, who is a single agent in a single laboratory.
To simplify the discussion, we suppose Alice has a master agent, $A_0$, with whom the $A_i$ communicate: $A_0$ may be $A_i$ for some $i$, or at a separate location. We assume the locations of Alice's laboratories are reliably known to
$A_0$ and that all Alice's agents (including $A_0$) have reliable clocks within their laboratories that are all synchronized. We also assume that Alice's laboratories are robust and secure enough that destruction/dislocation attacks on them are not a concern -- or at least, that it is a reasonable working assumption that such attacks will not succeed, perhaps because if they did then the loss would be so devastating that the failure to verify Bob's location
securely becomes irrelevant. However, Bob cannot directly verify his position: there is no trusted global GPS and any incoming reference data he might use could be
spoofed. His laboratory may also be technologically more limited than Alice's and may be vulnerable to destruction/dislocation attacks.

The $A_i$ are at separate sites that are also separated (and may be quite distant) from the location $L$, which lies within their convex hull. We can consider 1D or 2D position verification, which assume that Bob and the $A_i$ are constrained to lie on a given line or plane. In the 3D case, there is no such constraint, of course. In $d$ dimensions, we must have $M \geq (d+1)$ to satisfy the convex hull condition.
The  $A_i$ send classical or quantum signals at appropriate times. They receive classical or quantum signals from Bob and verify that these are the prescribed responses (up to a stipulated error rate) and are received at the correct times (up to stipulated timing errors).    

Following Ref. \cite{K11.1}, we additionally assume that $B$ and the $A_i$ can keep some classical information secure \edit{within} their respective laboratories, i.e. that adversaries cannot obtain this information unless and until the relevant agent chooses to transmit it outside their laboratory. We assume this holds true even if adversaries are able to move or destroy Bob's laboratory.  While this is a strong assumption, it is a standard one in quantum key distribution \edit{(QKD)} and other areas of cryptography.  It effectively counters replacement attacks, since if a tagging device within Bob's laboratory contains a significant amount of secret information, adversaries have only a small probability of constructing an identical replacement. 

\edit{More precisely}, there is only a small probability that \edit{{\it any given attempted replacement}} will be identical. Adversaries could construct a set of replacements $R_i$, where $R_i$ contains key $k_i$,
so that each possible key is contained within one of the replacements. This guarantees that the tag has been precisely replicated, but does not give the adversary a way of identifying which replacement is the replica, and does not facilitate a useful spoofing attack. \edit{Even for a relatively short key of 62 bits (as in the security analysis provided in section \ref{secV}), a successful attack of this form \edit{requires $2^{62}\approx 10^{18}$} replacements.}

\section{Scheme 1: Bob can keep a classical key secret but has no trusted clock}
  \label{secIII}

We work with the following slight variation of the scheme of
Ref. \cite{K11.1}. Let Alice have $M$ agents $A_1,\ldots, A_M$ surrounding the expected location $L$ of Bob.   We assume this expected location $L$ is initially known to Alice, and may not
necessarily be known to Bob. We assume that in the absence of adversarial interference $L$ is fixed throughout the position verification protocol, i.e. that Bob remains stationary.  

Alice and Bob agree on a sufficiently large integer $n$, and
on a sufficiently small tolerable error rate $0\leq \gamma <<1$, which act as security parameters. In this scheme we assume Bob does not possess a trusted clock.

We simplify the description by making the idealized assumption that all communications take place at or near the speed of light through vacuum, which we denote by $c$.  The scheme tolerates small delays, at the cost of reducing the precision to which Alice can verify Bob's precision. 

In essence, the scheme requires Alice and Bob to authenticate to each other using previously distributed secret keys. For all
$i=1,2,\ldots, M$, Alice's agent $A_i$ and $B$ perform the following actions.   

\begin{enumerate}
\item At some time prior to the position verification protocol, $A_i$ and $B$ share a secret key $k_i = \{ k_{i1} , \ldots , k_{iN} \}$, where each substring $k_{ij}$ has $n$ bits, and $N$ is as large as required. They keep this secret from eavesdroppers, although $A_i$ may share it securely with other $A_j$. The key sharing may be done arbitrarily far in advance. Let $k_{ij}=(q_{ij}, r_{ij})$, where $q_{ij}$ and $r_{ij}$ are sub-strings of $k_{ij}$ comprising the first $m$ bits and the last $(n-m)$ bits of $k_{ij}$, respectively.  Here `q' and `r' stand for `query' and `reply'. For simplicity, we assume the key-sharing is error-free, i.e. that
  $A_i$ and $B$ have identical versions of $k_i$. This can be
  achieved, with high probability, by standard key reconciliation methods. A small allowed key sharing error rate can also be incorporated in the error-tolerant ($\gamma >0$) versions of the position verification protocol, as described below. 

\item $A$ and $B$ may also, prior to the position verification
  protocol, share a separate secret key $k$. This key may be expanded as and when needed by \edit{QKD}, if $A$ and $B$ have the appropriate resources, and used to extend the secret keys $k_i$ (i.e. to
  increase the number of \edit{substrings $N$).} 

The position verification protocol typically involves many rounds of queries and replies. We describe here round $j$.

\item Each $A_i$ sends \edit{query-message-$ij$, comprising} the plain text `Query $i j$' followed by
  $q_{ij}$ (the query sub-string of $k_{ij}$) to $B$. The messages are sent at light speed, timed so that they should arrive at \edit{location $L$} at time $T_j$.    \edit{The messages are sent in such a way that, in the absence of interference, they can be distinguished even if they arrive simultaneously.   For example, they may be sent using different frequencies, or using a code that allows superimposed classical 
  messages to be distinguished.}

\item  \edit{$B$ processes the messages purportedly received from the various $A_i$ sequentially, using some ordering algorithm for distinguishable messages received simultaneously.   We thus, in the next step, describe his response to a single message in the sequence.}

\item  \edit{$B$ receives the message $(`\text{Query}~
  ij'$',\edit{$q'_{ij'})$}, purportedly from
  \edit{$A_i$}.}  Because of possible adversarial interference, we do not assume \edit{that $j'=j$, even if the message arrives at $T_j$, 
  nor that $A_i$ has in fact sent query-message-$ij$, nor that}  \edit{$q'_{ij'} = \edit{q_{ij'}}$}.
  \edit{$B$ keeps a register $R_i$ recording the second index $j$ of the last query-message-$ij$ that he received and validated with 
  first index $i$.} 
  \edit{If $R_i = (j'-1)$, then $B$ sets $R_i$ to $j'$}.   If not, he aborts the protocol (i.e. does not continue with the steps below)
  and stops responding to any future queries.
  (Here and below, we give the simplest response to detection of an apparent spoofing.  $B$ stops communicating; the $A_i$ become aware that the protocol has failed; they presumably take appropriate action.  Of course, other ways of proceeding are possible and in some circumstances preferable.   For example, $B$ may continue to respond to valid queries so long as the proportion of invalid queries is \edit{smaller than
  some pre-agreed threshold.  With suitable adjustments of security parameters, this modified protocol can continue to give useful security guarantees.})

  \item \label{bverify}   
  $B$ checks whether the Hamming distance \edit{$d(q'_{ij'}, q_{ij'}) \leq \gamma m$}. If so, he accepts the query as authentic.
  If not, he aborts, as above.  

  Timing: If $B$ is indeed at location $L$, the messages arrive at time $T_j$, and his authentication takes place within the time interval $[T_j,T_j +\delta_1]$, where $\delta_1$ is the processing time required.

\item \label{breply} If $B$ authenticates \edit{the query purporting to come from \edit{$A_{i}$}}  in the previous step, he sends the plain text \edit{`Reply \edit{$i j'$'}} followed \edit{by $r_{ij'}$} (i.e, the reply sub-string \edit{of $k_{ij'}$) to $A_{i}$}.  Otherwise $B$ does not respond, and accepts no subsequent queries \edit{from any $A_i$}, even if they are authenticated. 

\edit{Depending on which option is more technologically convenient, $B$ either broadcasts his response to all the $A_i$. 
or sends it just to the agent $A_{i}$ identified by the authenticated $\text{Query}~ij'$.  (Note that it is possible that $A_{i}$ is not the sender even though Bob has authenticated the query.)   We call these respectively the {\it broadcast} and {\it narrowcast} versions of the protocol.
In the broadcast version, if $B$ broadcasts simultaneous messages to more than one $A_i$, the messages are sent in such a way that, in the absence of interference, they can be distinguished even if they arrive simultaneously (as above).}

Timing: If $B$ is indeed at $L$, he sends his reply by the time $T_j+\delta_1+\delta_2$, where $\delta_2$ is the time taken by $B$ to transmit.

\item  \edit{In the narrowcast version, suppose that \edit{$A_{i}$} receives (`\text{Reply}~$i' j'$', \edit{$r_{i'j'}'$)}, purportedly
  from Bob.   
   If $i=i'$, and $j'=j$ (where $j$ is the current round, i.e. $A_i$
   has received and authenticated replies from all rounds $k<j$),
   $A_i$ verifies that the Hamming
   distance $d(r_{ij},r'_{ij})\leq \gamma (n-m)$ \edit{and that she has received the reply $r'_{ij}$ not later than the time $T_j +\frac{d_i}{c} +\delta_L$, where $\delta_L$ has been agreed in advance by Bob and Alice and must satisfy $\delta_L\geq \delta_1+\delta_2$.} If so, she accepts the reply as \edit{authentic and} sends the confirmation message `$s_{ij}=1$' via an authenticated channel to the master agent $A_0$.  Otherwise, she sends the message `$s_{ij}=0$'. } 

\edit{In the broadcast version, $A_i$ ignores all messages not of the form (`\text{Reply}~$i j'$', $r_{ij'}'$) for some string $r'$ and some $j'$.   That is, $A_i$ considers only replies apparently addressed to her.    For those messages, she follows the verification steps
for $j'$ and $r'$ above, and sends confirmation messages as above.  }

  Timing: If $B$ is indeed at $L$, then $A_i$ authenticates his reply by the time
$T_j+\frac{d_i}{c}+\delta_1+\delta_2+\delta_3$,
where $d_i$ is the distance from $L$ to $A_i$ and
$\delta_3$ is the time taken for $A_i$ to authenticate. She sends her confirmation message at time 
  $T_j+\frac{d_i}{c}+\delta_1+\delta_2+\delta_3+\delta_4$, where
  $\delta_4$ is the time taken by Alice to generate and transmit the confirmation message.

\item If $A_0$ receives $s_{ij}=1$ and authenticates it as a message
  from $A_i$ for all $i=1,2,\ldots,M$, then she
  authenticates that the location of $B$ at time $T_j$ was $L$, within
  some position uncertainty given by \edit{$c\delta_L$}. If she receives $s_{ij}=0$ for some $i$ then the
  position verification in round $j$ fails.  

  Timing:  Suppose that the authenticated channel from $A_i$ to $A_0$ had length $d'_i$ and that 
  messages travel on it at speed $c'_i \leq c$.    $A_0$   completes the authentication by the time
  $$T_A=T_j+\edit{\max_i \edit{\Bigl\{\frac{d_i}{c} + \frac{d'_i}{c'_i}\Bigr\}}}+\delta_1+\delta_2+\delta_3+\delta_4+\delta_5 \, , $$
\edit{where $\delta_5$ is the time taken for $A_0$ to compute whether all confirmation messages satisfy
  $s_{ij}=1$.}
\end{enumerate}

\textbf{Comments on Geometry:} A successful verification by $A_i$ guarantees
that\edit{, at time $T_j$, $B$ was} within a ball $B_i$ with $A_i$ at the centre\edit{, with radius $d_i+c\delta_L$}. 
Using the
information of her agents, $A_0$ can verify $B$'s location to \edit{have been} within the
intersection of the balls $B_1,\ldots,B_M$. \edit{In the ideal case $\delta_L=0$, this intersection is exactly the location $L$.} Thus, by increasing $M$,
and appropriately locating Alice's agents we can reduce the
uncertainty of $B$'s location in the verification scheme.

Note that a proof of principle implementation for 3D position verification could be carried out \edit{for} $M=2$, with $A_1$ and $A_2$ approximately collinear with $L$, on
opposite sides of $L$.
If the collinearity is a good approximation, this may verify location
to good precision.
The precision is greater if we may assume $A_1$, $A_2$ and $B$ are constrained to
lie exactly on a line, so that we effectively carry out position verification in 1D.

An implementation with $M=1$ can only guarantee that $B$ is within a ball surrounding Alice's
sole agent $A_1$.   Nonetheless, implementation of $M=1$ would
allow a proof of principle test of the technology, since we
can extrapolate the results to $M>1$ and estimate the precision that would be obtained \edit{for any
given configuration of the $A_i$ and $B$.} 

\edit{\textbf{Comment on key lengths and error tolerance:} It may be useful to vary the key
lengths in some scenarios, so that $k_{ij}$ has length $n_{ij}$. 
It may also be useful to allow the error thresholds to vary, 
so that communications in round $j$ from $A_i$ to $B$ have threshold 
$\gamma^{\rightarrow}_{ij}$, and from $B$ to $A_i$ have threshold 
$\gamma^{\leftarrow}_{ij}$.
For simplicity we consider fixed $n$ and $\gamma$ in our discussion.}

\section{Time delays of Scheme 1}
\label{secIV}
We identify three main types of time delays. 

\begin{enumerate}

\item $\Delta_L = \delta_1+\delta_2$ provides the uncertainty
  in $A_i$'s estimate of $d(A_i , L)$, assuming that she is confident
  of the value of $\delta_3$, which is determined by her own
  equipment.  (In practice there will presumably be at least slight
  \edit{uncertainties} in $\delta_3$.   We neglect these for simplicity;
  they can be included in the calculation of $\Delta_L$ if significant.)
  We would like to make $\Delta_L$ as short as possible in order for
  the location $L$ to be verified as precisely as possible. \edit{As mentioned above, we require $\Delta_L\leq \delta_L$.}

\item $\Delta_V = \delta_3 + \delta_4+\delta_5$ comprises the delay
  that $A_0$ takes in learning $A_i$'s estimate $d(A_i , B)$, after 
  $A_i$ receives $B$'s response.   During the interval $\Delta_L + \Delta_V$, $B$ could be
  displaced, with $A_0$ learning this (if at all) only later.  
  So we would also like to make $\Delta_V$ as short as possible,
  all else being equal.   

\item \label{III.3} $\Delta_R$ denotes the time difference between one
  verification and the following one, where $R$ denotes
  `repetition'. All else being equal, we would like to repeat the
  location verification protocol as frequently as possible and thus
  minimize $\Delta_R$.   
\end{enumerate}

The relative importance of minimizing $\Delta_L , \Delta_V$ and
$\Delta_R$, and the value of tradeoffs among them, depends on
the scenario.  One factor is whether Bob's potential displacement
(either by wandering or by the action of adversaries)
is bounded only by $c$, or whether in practice 
a significantly lower bound (such as the speed of sound in air,
or the speed of the fastest planes currently available) is \edit{justified.}

\section{Some possible attacks on Scheme 1}
\label{secV}

\edit{
Our security analysis below applies for arbitrarily powerful spoofers, who might have arbitrarily advanced quantum technology and who could share an arbitrarily large amount of entanglement. Our analysis is based on the assumption that Bob can keep classical information secret from {\edit spoofers}.   \edit{The assumption that collaborating parties have laboratories in which they can keep data secure is 
standard (and necessary) for quantum key distribution schemes and many other quantum cryptographic protocols.
We believe it is equally reasonable in many scenarios in which position verification is required.  
Nonetheless, it is ultimately a technological assumption, whose validity should be
examined in any given application and scenario.   
}}

\subsection{Desynchronizing Alice's clocks}

As mentioned above, Alice's agents must keep their laboratories
securely synchronized to a common reference frame $F$ during the verification
scheme.  \edit{We assume they trust their locations.   This is reasonable
in scenarios in which Alice has a secure and stable infrastructure.   
They also require} secure clock synchronization.
This is a general issue in relativistic quantum cryptography.
\edit{In practice, it} is often partially addressed by keeping clocks synchronized
using GPS devices within the required time
uncertainty
\cite{LKBHTKGWZ13,LCCLWCLLSLZZCPZCP14,LKBHTWZ15,VMHBBZ16,ABCDHSYZ21}.
However, adversaries may spoof the GPS signals \edit{\cite{TPRC11}}, \edit{desynchronizing}
Alice's clocks.

To defend against such attacks, the $A_i$  may initially synchronize
their clocks in
a single secure laboratory (for example $A_0$'s) and then displace
these securely to their
separate laboratories.   To counter clock drift, this process
could in principle be repeated
at suitably short intervals during the protocol, ensuring
that the $A_i$'s clocks are repeatedly re-synchronized
with new incoming synchronized clocks.
Note that this requires not only a supply of accurate clocks,
but also secure distribution channels that  
ensure the clocks remain \edit{very well synchronized}
as they are distributed.   

\subsection{Impersonating Alice and Bob}

The spoofer may have multiple agents $S_k$ at
separate locations: we refer to them collectively as $S$.  
The spoofer $S$ could try to impersonate $B$ in an strategy that combines spoofing $A_i$ to $B$,
with the aim of learning the strings $r_{ij}$, and spoofing $B$ to $A_i$.
We consider this for a single $A_i$; the discussion obviously extends to
multiple agents.   

Suppose that rounds up to $(j-1)$ have been honestly completed
by $A_i$ and $B$.   Before $B$ receives $A_i$'s communication for
round $j$, $S$ could send the plain text `Query $i j$' to $B$ followed by some
string $q_{ij}'$ of $m$ bits.  Assuming $S$ has no 
previous information about $q_{ij}$, with probability $2^{-m}$,
$q_{ij}'=q_{ij}$ and $B$ sends $r_{ij}$ to $S$.  $S$ can then send
$r_{ij}$ to $A_i$, impersonating $B$ when $A_i$ later
sends the authentic  `Query $i j$' with string $q_{ij}$.   

With probability $1-2^{-m}$, $q_{ij}'\neq q_{ij}$, in which case $B$
does not transmit $r_{ij}$ in following queries by $S$ or Alice.  In
this case, $S$ generates a random guess $r_{ij}'$ \edit{of} $r_{ij}$ and
sends it to $A_i$ after the plain text `Reply $i j$', assuming she
does not have any previous knowledge about $r_{ij}$.  $S$ succeeds in
this case with probability $2^{-(n-m)}$.

Thus, if the keys $k_{ij}$ are perfectly random and secret, $S$'s
probability to succeed in impersonating $B$ in the previous strategy
is
\begin{equation}
\label{Eve}
\edit{P_\text{S}}=\Bigl(\frac{1}{2}\Bigr)^m+\Biggl[1-\Bigl(\frac{1}{2}\Bigr)^m\Biggr]\Bigl(\frac{1}{2}\Bigr)^{n-m}.
\end{equation}

We have assumed here that $\gamma=0$, i.e., neither $A$ nor $B$ accept
errors in the received strings.
$P_\text{$S$}$ is minimized (for fixed even $n$) for $m=\frac{n}{2}$,
when,
\begin{equation}
\label{spoof2}
P_\text{$S$}=2\Bigl(\frac{1}{2}\Bigr)^{\frac{n}{2}}-\Bigl(\frac{1}{2}\Bigr)^{n} \edit{\leq } 2^{1 - \frac{n}{2}} \, . 
\end{equation}

This gives a strong security bound of $P_\text{$S$} \approx 10^{-9}$
with relatively short keys of $n=62$ bits.

\edit{
We now consider the case $\gamma>0$. We have
\begin{equation}
\label{spoof2.1}
P_\text{$S$}(\gamma)=\Bigl(\frac{1}{2}\Bigr)^m\lvert Q_m^\gamma\rvert+\Biggl[1-\Bigl(\frac{1}{2}\Bigr)^m\lvert Q_m^\gamma\rvert\Biggr]\Bigl(\frac{1}{2}\Bigr)^{n-m}\lvert Q_{n-m}^\gamma\rvert, 
\end{equation}
where $Q_N^\gamma=\{ x\in\{0,1\}^N\vert w(x)\leq\gamma N\}$, and where $w(x)$ denotes the Hamming weight of the bit string $x$.

We assume that $n$ is even and that $m=\frac{n}{2}$. We obtain from (\ref{spoof2.1}) that
\begin{eqnarray}
\label{spoof2.2}
P_\text{$S$}(\gamma)&=&2\Bigl(\frac{1}{2}\Bigr)^m\lvert Q_m^\gamma\rvert-\Bigl(\frac{1}{2}\Bigr)^{2m}\bigl(\lvert Q_m^\gamma\rvert\bigr)^2\nonumber\\
&\leq&2^{(1-m)}\lvert Q_m^\gamma\rvert\nonumber\\
&\leq&2^{(1-m)}2^{mh(\gamma)}\nonumber\\
&=&2^{\bigl[1-m(1-h(\gamma))\bigr]},
\end{eqnarray}
where in the third line we used that $\lvert Q_m^\gamma\rvert\leq2^{mh(\gamma)}$, which is shown
in Sec. 1.4 of Ref. \cite{Lintbook}, and where $h(\gamma) = -\gamma \log_2 \gamma - (1 -
\gamma ) \log_2(1 - \gamma )$ is the binary entropy of $\gamma$.

}


For example, even with a high error tolerance, $\gamma = 0.05$ \edit{(giving $h(0.05)=0.2864$) and taking $n=88$, we obtain from (\ref{spoof2.2}) that $P_\text{$S$}(0.05) < 10^{-9}$.}


We have assumed the keys are perfectly random, i.e.,
that $S$ has zero information about them. 
In practice this will not be quite correct,
but keys can be made close enough to random to make
corrections negligible.

\subsection{Obtaining the keys}

$S$ could try to learn the keys $k_{ij}$ before they are used, i.e.,
before $B$ sends the $k_{ij}$'s to Alice's agents. But by assumption the scheme is
secure against these attacks: we assume Alice and Bob generate and distribute
the keys secretly, and store them secretly until Bob communicates the
secret keys.

\subsection{Adding time delays}

$S$ can add time delays in any communications between any of Alice's
agents and $B$. This can increase the uncertainty in $B$'s location
authenticated by Alice and/or delay Alice's verification. If $S$ thereby causes the
time delays between the $A_i$ and $B$ to be larger than
acceptable thresholds, or for the communications in one or both
directions
to become out of sequence, then this causes the position verification to fail.
We assume in this case Alice responds, for example by inspecting $B$'s location
with other physical means.

A limiting case is that $S$ can jam the communications altogether. 
These delay or jamming attacks are unavoidable in practice unless the communication channels
cannot be accessed by $S$.
Inaccessible private channels is a strong assumption
not generally made in quantum cryptography.
One practical reason not to make this assumption 
is that it requires secure laboratories linked by a network of securely hardened channels.
In position verification applications, these channels might typically be
\edit{$\sim 10-10^5$km long} (ranging from small scale networks on Earth to
high Earth orbit satellites) and would need to be approximately straight
line. 
A theoretical reason not to make it is that it trivialises position
verification (PV) as a task.    Given inaccessible private 
channels, PV can be securely implemented simply by
exchanging messages, without using secret keys or any quantum
communications, provided the
channel transmission times are reliably known.   

If $S$ adds suitably short time delays, $A$ can verify $B$'s location
within tolerable error bounds.   If she adds longer time delays, she prevents
$A$ from verifying $B$'s position, but alerts \edit{$A$} to her
interference.  So the protocol is secure (in the sense claimed)
against delay or jamming attacks.   

\edit{\subsection{Altering Bob's records}

If $S$ can, without detection, alter Bob's record of whether previous queries were authenticated, she can send 
repeated queries of the form query-message-$ij$, altering the record so that Bob has no record 
of each failed authentication.   Bob will thus continue the protocol after failed authentication.
$S$ can thus continue until she successfully guesses the query string $q_{ij}$, and Bob's response
will provide her with $r_{ij}$.    If she is able thus to obtain $r_{ij}$ before the 
authentic query-message-$ij$ is sent, she can spoof a response to this challenge.  If she
is able to do this for all $ij$, she can systematically and indefinitely mislead the $A_i$
as to $B$'s location.    

Timing constraints may restrict the scope of this attack, since as described it requires $S$
to make $\sim 2^m$ guesses at $q_{ij}$ to obtain $r_{ij}$.    However, if $S$ is also able
to alter the data in Bob's register $R_i$, she can set it so that Bob will accept a 
query-message-$ij$ for some value of $j$ that may not authentically be sent until 
some (perhaps far) future time.   This enlarges the time window during which she can
send guesses at query-message-$ij$. 

Also, if $S$ can, without detection, alter Bob's records of the $q_{ij}$, in a way that
allows her to choose the altered string (although not read the original string), 
she can create new query keys $q_{ij}^S$ that she knows.
If she is able to do this before the authentic query-message-$ij$ is sent, she can use $q_{ij}^S$ to 
send a spoof query-message-$ij$, obtain the response $_{ij}$, and use this to spoof responses to 
authentic queries.    If she
is able to do this for all $ij$, she can systematically and indefinitely mislead the $A_i$
as to $B$'s location.

The protocol thus requires that $B$ can keep classical data secure
against alteration, as well as keeping the key strings private (i.e. secure against reading).
The ability to ensure that classical data within a secured site is unalterable is also a 
standard cryptographic assumption.    However it is worth noting that it neither necessarily implies
nor is necessarily implied by data privacy.   
}

\section{Quantitative considerations for scheme 1}
\label{secVI}
We consider for simplicity a 1D implementation in which Alice has two agents (case
$M=2$), $A_1$ and $A_2$. Let $A_1$, $A_2$ and $B$ be on the same line,
with $B$ between Alice's agents at equal distance from each of them.

We assume that the communication channel between $A_1$ ($A_2$) and $B$ is optical and in free space,
transmitting at the speed of light $c$.

\subsection{Bob performs information processing with electronic circuits}

Steps \ref{bverify} and \ref{breply} comprise $B$ receiving Alice's query signal encoded in
\edit{light or other electromagnetic signals}, converting these to electronic signals, authenticating the
request originated from Alice, and then encoding the reply in \edit{light or other electromagnetic signals}
and transmitting to $A_i$, for $i=1,2$. In practice, a circuit
comprising FPGAs could be used to
perform these computations as fast as possible. \edit{Note that the signals exchanged between Alice and Bob \edit{in the position verification protocol are classical.   Hence they} can be sufficiently intense to deal with losses and errors. This is a \edit{significant} practical advantage compared to position authentication schemes that need to transmit quantum states between Alice and Bob (e.g., \cite{patent, malaney2010location,chandran2009position,BCS22,ES23})}.

As an illustration, we take \edit{$n \sim 62$}, with $\gamma = 0$. \edit{We note that because our scheme only involves classical communication and classical processing, it is sensible to assume zero errors, in contrast to schemes that require quantum communication or \edit{quantum information processing}.} 
An FPGA simply needs to compare received and stored key strings. 
Ref. \cite{VMHBBZ16} performed in
2016 completed a round including more complex computations and
communication between adjacent FPGAs with a string of 128 bits in 1.8
$\mu$s.   With these devices, assuming processing time is
approximately
linear in string length, our simpler computation with \edit{$n=62$}
bits should be completed within  \edit{$\leq 0.88 \mu$s}, giving
uncertainty \edit{of $\leq 264$m} in $B$'s location.  

With state of the art FPGAs, we estimate that
$B$'s verification might be completed within $\sim 10$ns,
giving an uncertainty of $\sim 3$m in $B$'s location.

If verification rounds take place every $\mu$s, the light
speed signalling bound implies that the tag can move
$\leq 300$m between rounds.  For $4$ verifiers, this
round frequency consumes \edit{$\approx 4 \times 124 \times 10^6 \approx 5
\times 10^8$} key bits per second.

If we assume, perhaps plausibly in many scenarios, that the
tag will not move faster than the speed of sound (in air at
sea level) from its expected location, verification rounds every $\mu$s mean that the
tag cannot move more than $\approx 3 \times 10^{-4}$m between
rounds.  With rounds every $ms$, this becomes $\approx 3 \times
10^{-1}m$.
This round frequency consumes \edit{$\approx 5 \times 10^5$} key bits
per second.   While still demanding, these resource requirements
seem achievable with present technology, and give (modulo
assumptions) good enough location precision to be useful
in many scenarios.  

\section{Scheme 2: Bob has a trusted synchronized clock}
\label{secVII}
Our second scheme introduces the assumption
that $B$ possesses a clock synchronised with Alice's clocks.
This allows Alice to specify a
signalling schedule in advance so $B$ can transmit at specified times.
Of course, it makes extra technological demands on $B$ and on the
size and security of his laboratory.   

The main difference between Schemes 1 and 2, is that Scheme 2
does not require Alice to send a query signal every time she wishes
to verify $B$'s location. Instead, she shares an authenticated
signalling schedule with $B$ in advance, with $B$ relying on his
synchronised clock to follow the schedule. This removes the need for
$B$ to authenticate each individual query from Alice in real time,
as he can authenticate the whole signalling schedule in
advance. This removes one source of delay in the protocol, namely
$B$'s authentication time, denoted as $\delta_1$ in the description
of Scheme 1.  It also potentially removes a second source of
delay, $B$'s transmission time $\delta_2$, since if Bob knows
$\delta_2$ he can adjust for it by starting the transmission
so that it {\it completes} (rather than starts) at the time $T_j$
stipulated for his round $j$ communication in $A$'s schedule.
More precisely, it potentially replaces $\delta_2$ by the
uncertainty $\Delta_2$ in Bob's transmission time, and
typically we expect $\Delta_2 \ll \delta_2$.
We assume this adjustment below.  

(We consider schemes involving only classical communication
between $A$ and $B$ here, but it is worth noting that removing these
delays is potentially even more valuable for schemes involving quantum
communications from $A$ and $B$ and quantum measurement and/or
information processing by $B$, since the latter steps are potentially
significantly slower than their classical counterparts.
It thus seems potentially advantageous, in scenarios in which it is
justifiable, also to allow $B$ and $A$
to share synchronized clocks in such schemes.
However, the advantage is lost if $B$ is not able to keep
information secret, since storing information for later
use exposes it to spoofers.   And if $B$ is able to keep
secret, the protocols discussed here using classical queries
and responses may generally be more efficient.
So this option may perhaps be useful only in the restricted
scenario where $B$ is able to keep quantum information secret
but not classical information.)

Scheme $2$ again assumes that Alice has $M$ agents $A_1,\ldots,
A_M$ surrounding the location $L$ of $B$. Alice and $B$ agree on a
sufficiently large integer $n$, and on a sufficiently small error rate
$0\leq \gamma <<1$, which act as security parameters. We also assume
that all communications take place at the speed of light through
vacuum, which we denote by $c$. For all $i=1,2,\ldots, M$, Alice's
agent $A_i$ and $B$ perform the following actions.

\begin{enumerate}

\item $A$ and $B$ share a secret key $k$ that is used to
  authenticate communications and may also be used as a
  one-time pad to keep communications secret.   This key
  may be expanded as and when needed by \edit{QKD}.  

\item At some time prior to the position verification protocol, $A_i$
  and $B$ share a secret key $k_i = \{ r_{i1} , \ldots , r_{iN} \}$,
  where each substring $r_{ij}$ has $n$ bits, and $N$ is as large as required.
  They keep this secret from \edit{spoofers}, although $A_i$ may share it securely with other $A_j$.
  The key sharing may be done arbitrarily far in advance.
  For simplicity, we assume the key-sharing is error-free, i.e. that
  $A_i$ and $B$ have identical versions of $k_i$.  This can be
  achieved, with high probability, by standard key reconciliation
  methods.   A small allowed key sharing error rate can also be
  incorporated in the error-tolerant ($\gamma >0$) versions of
  the position verification protocol, as described below.
  
\item $A$ sends an authenticated message to $B$,
  specifying the times at which he is required to verify his
  location. This can be done arbitrarily far in advance of step 3, and
  possibly after or concurrent with step 2.  In the simplest version,
  this message is public.   Alternatively, it could be encrypted,
  to prevent $S$ from learning the verification schedule.

\item If $T_j$ is the $j$-th time $B$ is required to verify his
  location, then \edit{to each $A_i$}
  he sends the plain text `Reply $i j$' followed by
  $r_{ij}$ , completing his transmission at time
  $T_j$, according to his clock.   \edit{As in Scheme 1, $B$ may either
  broadcast or narrowcast his replies.  We describe the narrowcast 
  version below; the minor modifications required for the broadcast
  version are as for Scheme 1.}

  Timing: this takes place by the time $T_j+\delta_d+\Delta_2$, where
  $\Delta_2$ is the uncertainty in the time it takes
  $B$ to transmit (consistently defined as in Scheme 1) and $\delta_d$
  is the time difference between Alice and $B$'s clocks, which may
  be non-zero if they are not perfectly synchronised.
  \edit{Note that we cannot assume $\delta_d >0$.
  We assume that $A_i$ and $B$ are confident that their clock technology will ensure, in the absence of adversarial attacks
  on their clocks, there is some bound $\delta_d^{\rm max} > 0$ such that $ |\delta_d | \leq \delta_d^{\rm max}$.
  In practice, this bound will be time-dependent.  For simplicity here we consider a single bound that is valid 
  throughout the duration of the protocol.}

\item $A_i$ receives (`\text{Reply}~$i' j'$', \edit{$r_{i'j'}'$}).   She
  verifies that $i'=i$, that $j'=j$ (where $j$ is the current round,
  i.e. $A_i$    has received and authenticated replies from all rounds $k<j$)\edit{, that}
  the Hamming distance $d(r_{ij}',r_{ij})\leq \gamma (n-m)$ \edit{and that she has received the reply $r'_{ij}$ not later than the time $T_j +\frac{d_i}{c} +\tilde{\delta}_L$, where $\tilde{\delta}_L$ has been agreed in advance by Bob and Alice, and which must satisfy $\tilde{\delta}_L\geq \edit{\delta_d^{\rm max}}+\Delta_2$}.   If so, she sends
  the confirmation message `$s_{ij}=1$' \edit{together with the time that she received $r'_{ij}$} to $A_0$.
  Otherwise she sends the message `$s_{ij}=0$'.

  Timing: If $B$ is indeed at $L$, then $A_i$ authenticates his reply by the time
$T_j+\frac{d_i}{c}+\delta_d+\Delta_2+\delta_3$,
where $d_i$ is the distance from $L$ to $A_i$ and
$\delta_3$ is the time taken for $A_i$ to authenticate. 
She sends her confirmation message at time 
  $T_j+\frac{d_i}{c}+\delta_d+\Delta_2+\delta_3+\delta_4$, where
  $\delta_4$ is the time taken by Alice to generate and transmit the confirmation message.

\item If $A_0$ receives $s_{ij}=1$ and authenticates it as a message
  from $A_i$ for all $i=1,2,\ldots,M$, then she authenticates that the
  location of $B$ was $L$, within some uncertainty given by 
  $c (\tilde{\delta_L} - \delta_d^{\rm max})$, at some time $T'_j$, where $| T'_j - T_j | \leq \delta_d^{\rm max}$. 
  This implies that the 
  location of $B$ at time $T_j$ was $L$, within some uncertainty given
  by \edit{$c\tilde{\delta}_L$}.   

 \edit{Timing:  Suppose that the authenticated channel from $A_i$ to $A_0$ had length $d'_i$ and that 
  messages travel on it at speed $c'_i \leq c$.    $A_0$   completes the authentication by the time
  $$T_A=T_j+\edit{\max_i \edit{\Bigl\{\frac{d_i}{c} + \frac{d'_i}{c'_i}\Bigr\}}}+\edit{\delta_d + \Delta_2} +\delta_3+\delta_4+\delta_5 \, , $$
where $\delta_5$ is the time taken for $A_0$ to compute whether all confirmation messages satisfy
  $s_{ij}=1$.}

\end{enumerate}

\section{Further attacks on Scheme 2}

\edit{
As in section \ref{secV}, our security analysis here applies for spoofers who \edit{may have arbitrarily advanced quantum technology and share an arbitrarily large amount of entanglement}, given our assumption that Bob can keep classical information \edit{secure} from the spoofers.
}

\subsection{Acting outside schedule}

$S$ may gain knowledge of the signalling schedule, potentially
providing her with information about $\Delta R$, the time difference
between subsequent location requests. This exacerbates the problem
illustrated in point \ref{III.3} in section III, as $S$ would know
when she is able to move $B$ (or encourage him to move, by spoofed GPS
or other means) without risking a location check. This
risk could be mitigated either by keeping the schedule suitably secret
(by sending it encrypted) and unpredictable, or by ensuring the scheduled times are frequent
enough that $S$ could not move $B$ a significant distance before the
next verification is due.   Note that the first option consumes a potentially
large amount of shared secret key \edit{and the second requires a potentially 
high key refresh rate}.

\subsection{Clock desynchronization}

Bob's clock may become desynchronized, either naturally or as a result
of $S$'s interference, introducing a time difference $\delta_d$
between Alice and Bob's clocks.
At least theoretically, this is a significant concern: if $S$ is able
to move $B$ physically at speeds arbitrarily close to $c$, she can cause his clock
to run arbitrarily slowly with respect to $A$'s lab frame.
However, slowing $B$'s clock delays his responses, which means that
$A$ will not incorrectly verify his position as guaranteed to be
close to $L$.
In principle, $S$ could also desynchronize $B$'s clock by altering
the gravitational field in his vicinity.   Such attacks are usually
ignored in relativistic cryptography, given that it is impractical
to create any significant effect.   In most applications of
relativistic cryptography, security is threatened only if the
effect is large enough that points believed by one party to be spacelike
separated are in fact timelike separated.
In this case, though, any degree of desynchronization affects the precision
of the position verification.   However, in practice any effect
seems likely negligible compared to other \edit{uncertainties}.

$A$ and $B$ might attempt to counter accidental or deliberate
desynchronization by exchanging authenticated messages to
keep their clocks synchronized.  This is complicated by the
fact that a simple synchronisation protocol requires knowledge
of $B$'s position, or at least his distance from the relevant $A_i$.
Synchronization schemes involving exchanges with several $A_i$ in
parallel still appear useful, but we will not pursue their analysis
here.

\section{Conclusion}
\edit{We have presented schemes for position verification in which the position verification queries and 
responses are purely classical, involving no quantum communication or quantum information processing.  
Quantum information transmission and measurement is required only to refresh the key via QKD.
Our schemes are practical to implement with current technology. Their security is based on a standard assumption in quantum cryptography, also made in QKD, that a classical key can be stored securely, as initially proposed in Ref. \cite{K11.1}. 

When QPV schemes use position verification queries and/or responses that involve quantum communications, they typically use photons to encode quantum states.   This poses challenges, including errors in state preparation, processing and measurement, losses, and security problems due to imperfect single-photon sources and single-photon detectors (e.g., photon-number splitting attacks \cite{HIGM95,BLMS00} and multiphoton attacks \cite{BCDKPG21}) and side-channel attacks (e.g., \cite{BCDKPG21}). The problem of losses is particularly challenging in schemes with large distances between the tagging device and the verifiers. 
An advantage of our schemes is that the queries and responses are purely classical.   Quantum communications are needed only
to replenish the key via QKD, which is secure against errors and losses.
Moreover, the QKD communications, unlike the position verification queries and responses, are 
not tightly time constrained.}

\edit{Given our assumptions, our schemes are secure against arbitrarily powerful quantum spoofers, who may share an arbitrary amount of entanglement. 
This is also an advantage compared to the best known quantum schemes, which have only been proved secure against spoofers that share an
amount of entanglement linear in the classical information) \cite{BCS22,ES23}.}

\acknowledgments We thank Daniel Oi\edit{, Jasminder S. Sidhu} and Siddharth Joshi for helpful
discussions. We  acknowledge financial support from the
UK Quantum Communications Hub grant no. 
EP/T001011/1.  G.C. is supported by a studentship from the Engineering
and Physical Sciences Research Council.
A.K. and D.P.-G. were supported in part by Perimeter Institute for
Theoretical Physics. Research at Perimeter Institute is supported by
the Government of Canada through the Department of Innovation, Science
and Economic Development and by the Province
of Ontario through the Ministry of Research, Innovation and Science.

\edit{
\textbf{Author contributions.} A.K conceived the project. A.K and D.P.-G. did the majority of the theoretical work,
with input from G.C. A.K. and
D.P.-G. wrote the manuscript with input from G.C.
}

\bibliography{quantumtaggingbiblio}
\end{document}